\documentstyle[aps,prl,preprint,floats,epsfig]{revtex}

\providecommand{\KS}{\mbox{${\mathrm K_{\mathrm S}^0}$}}

\providecommand{\PZ}{\mbox{$\pi^0$}}

\providecommand{\BZ}{\mbox{${\mathrm B}^0_{\mathrm d}$}}

\providecommand{\DZ}{\mbox{${\mathrm D}^0$}}
\providecommand{\DZBAR}{\mbox{$\overline{{\mathrm D}^0}$}}

\providecommand{\DSTP}{\mbox{${\mathrm D}^{\ast+}$}}

\providecommand{\JPSI}{\mbox{$\mathrm{J}/\psi$}}

\providecommand{\slowpi}{\mbox{$\pi_{\rm slow}^+$}}
\providecommand{\CPV}{{$CP$ violation}}
\providecommand{\ACP}{\mbox{${\cal A}$}}

\providecommand{\QSIG}{\mbox{$Q\in [3.3,8.3]\ {\rm MeV}$}}
\providecommand{\OLDKSPZ}{\mbox{$(-1.8 \pm 3.0)\%$}}
\providecommand{\KP}{\mbox{\KS\PZ}}
\providecommand{\KK}{\mbox{\KS\KS}}
\providecommand{\PP}{\mbox{\PZ\PZ}}
\providecommand{\SLOWPI}{\mbox{$\slowpi$}}

\providecommand{\RAWKK}{\mbox{$(-14\pm 14)\%$}}
\providecommand{\RAWPP}{\mbox{$(+0.1\pm  4.8)\%$}}
\providecommand{\RAWKP}{\mbox{$(+0.0\pm  1.1)\%$}}
\providecommand{\AKK}{\mbox{$(-23\pm 19)\%$}}
\providecommand{\APP}{\mbox{$(+0.1\pm  4.8)\%$}}
\providecommand{\AKP}{\mbox{$(+0.1\pm  1.3)\%$}}

\begin{document}

\preprint{\tighten\vbox{\hbox{\hfil CLNS 00/1708}
                        \hbox{\hfil CLEO 00-27}
}}

\title{Search for {%\mathversion{bold} 
$CP$ violation in 
$\DZ\to\KS\PZ$, $\DZ\to\PP$ and $\DZ\to\KK$ decays}
}  

% Your author list ***DOES NOT*** go here!
% is goes below where you are instructed to insert it...
\author{CLEO Collaboration}
\date{\today}

\maketitle
\tighten

\begin{abstract} 
% Insert abstract here.
 We have searched for $CP$-violating asymmetries in neutral
charm meson decays in 13.7 fb$^{-1}$ of $e^+e^-$ collision data
at $\sqrt{s} \approx 10.6\ {\rm GeV}$ with the CLEO detector.
The measured asymmetries in the rate of \DZ\ and \DZBAR\ decays to
\KP, \PP\ and \KK\ final states are
\AKP, \APP\ and \AKK, respectively.

\end{abstract}
\newpage

{
\renewcommand{\thefootnote}{\fnsymbol{footnote}}

% Insert author and address list here
\begin{center}
G.~Bonvicini,$^{1}$ D.~Cinabro,$^{1}$ M.~Dubrovin,$^{1}$
S.~McGee,$^{1}$ G.~J.~Zhou,$^{1}$
A.~Bornheim,$^{2}$ E.~Lipeles,$^{2}$ S.~P.~Pappas,$^{2}$
M.~Schmidtler,$^{2}$ A.~Shapiro,$^{2}$
% ...
\end{center}
 
\small
\begin{center}
$^{1}${Wayne State University, Detroit, Michigan 48202}\\
$^{2}${California Institute of Technology, Pasadena, California 91125}\\
% ...
\end{center}

\setcounter{footnote}{0}
}
\newpage

% Insert body of the text here.
 Measurable $CP$ violating phenomena in strange~\cite{KTeV,NA48} and
beauty~\cite{BaBar,Belle,CDF} mesons are the impetus for numerous
current and future experiments~\cite{KOPIO,KAMI,BTeV,LHCb} that
are expected to challenge the standard model (SM) description of the
weak interaction. In contrast the SM predictions for $CP$ violation %(\CPV) 
in the charm meson system of
${\cal O}(0.1\%)$~\cite{Bucella} are probably not attainable by current
experiments, although recent conjectures~\cite{Bigi} 
indicate that direct $CP$ 
violating effects may be as large as
${\cal O}(1\%)$. 
Thus an observation of \CPV\  in charm decays %v2
exceeding the percent level would be strong evidence for non-SM %v2
processes. %v2

 Previous searches for mixing-induced~\cite{cleo.dmix} or direct
\CPV~\cite{cleo.cpv,pdg} in the neutral charm meson system have
set limits of $\sim 30\%$ or a few percent, respectively. We present
results of searches for direct \CPV\ in neutral charm meson decays
to pairs of light pseudoscalar mesons: \KS\PZ, \PZ\PZ\ and \KS\KS. Decays
to the latter two final states are Cabibbo-suppressed, thus enhancing
the possibility that interference with non-SM amplitudes could produce
direct \CPV. A previous search by CLEO~\cite{cleo.cpv} for direct \CPV\ in \DZ\ and \DZBAR\ decays
to \KS\PZ\ in 2.7 fb$^{-1}$ of $e^+e^-$\  collision data established
$\ACP(\KS\PZ) = \OLDKSPZ$ with the definition
\begin{equation}\label{eqnA}
\ACP(f) \equiv \frac{\Gamma(\DZ\to f)-\Gamma(\DZBAR\to f)}
                    {\Gamma(\DZ\to f)+\Gamma(\DZBAR\to f)}
\end{equation}
\noindent where $f$ is the final state.

 The current results are based upon 13.7 fb$^{-1}$ of $e^+e^-$\  collision
data at $\sqrt{s} \sim 10.6 \ {\rm GeV}$ accumulated with two configurations
of the CLEO experiment at the Cornell Electron Storage Ring (CESR).
Approximately one-third of the data were accumulated with the CLEO II
configuration~\cite{cleoii} that consists of 3 nested cylindrical wire
chambers surrounded by a CsI(Tl) electromagnetic calorimeter immersed 
in a 1.5T solenoidal magnetic field. The 3.5 cm radius beam pipe and
innermost wire chamber were replaced by a 2 cm radius beam pipe 
and a three-layer
double-sided silicon vertex detector in CLEO II.V~\cite{cleoiiv}. 
In addition the gas mixture in the main drift chamber was changed from an
argon:ethane to helium:propane~\cite{gas} mixture for improved
charged particle momentum resolution and efficiency.

 The Monte Carlo simulation of the CLEO II and CLEO II.V detector configurations
was based upon GEANT~\cite{geant}. Simulated events were processed in the
same manner as the data. We used a simulated sample of $e^+e^-\to q\bar{q}$
($q=u,c,s,d$) events representing a luminosity comparable to that of
the data to determine selection criteria and investigate 
some  % v2
systematic effects.
Systematic uncertainties in the asymmetry measurements are determined % v2
from the data when possible. % v2

  The charge of the slow pion produced in the 
decay $\DSTP\to\DZ\slowpi$ identifies
the flavor of the neutral charm meson at production (charge conjugation
is implied throughout unless explicitly stated otherwise). Candidate $\slowpi$
must be well-reconstructed tracks originating from a cylinder of radius 3 mm
and half-length 5 cm centered on the $e^+e^-$\  interaction point. A minimum
momentum requirement on \DZ\ candidates of 2 GeV$/c$ sets a lower limit
on the \SLOWPI\ momentum of approximately 95 MeV$/c$.

 Candidates for the decay $\KS\to \pi^+\pi^-$ are formed from opposite-sign
pairs of charged particles within 8 (9) MeV of the known \KS\ mass~\cite{pdg}
for the \KS\KS\ (\KS\PZ) final state. The reconstructed \KS\ decay vertex 
must be separated from the interaction point by at 
least 3 standard deviations ($\sigma$) where $\sigma$
is calculated from the track covariance matrices. 
In addition the $\chi^2$ for each
\KS\ daughter track to originate from the interaction point is required to
be larger than 2.5. The latter two requirements not only suppress combinatorial
background arising from random combinations of $\pi^+\pi^-$ candidates but 
also diminishes
the contribution of $\DZ\to h^+h^-\KS$ and $\DZ\to h^+h^-\PZ$ ($h=\pi,K$) 
backgrounds to the 
\KS\KS\ and \KS\PZ\ final states, respectively.

 Neutral pion candidates are formed from pairs of electromagnetic showers
in the CsI(Tl) calorimeter unassociated with a charged track. Showers in the
barrel (end cap) region of the calorimeter must exceed 30 (50) MeV 
to be considered
as a \PZ\ daughter candidate where the barrel is the region of the calorimeter
at least $45^\circ$ from the $e^+e^-$\  collision axis. 
The invariant mass of \PZ\ candidates
must lie within 20 (18) MeV of the known \PZ\ mass~\cite{pdg}
for the \PP\ (\KP) final state.

 Neutral pion and \KS\ candidates are kinematically constrained to the \PZ\ and
\KS\ mass~\cite{pdg} and combined to form \DZ\ candidates. The mass constraints 
on the \DZ\ 
daughter candidates improves the \DZ\ mass resolution by 8\%, 5\% and 17\%
 for the
\KS\PZ, \PZ\PZ\ and \KS\KS\ final state, respectively. A final requirement 
is placed
on $\cos\theta_d$ where $\theta_d$ is the angle in the \DZ\ rest frame between
the \PZ\ (\KS) direction and the \DZ\ flight direction 
for \PZ\PZ\ (\KS\PZ\ and \KS\KS)
decays. Combinatorial background due to low momentum \PZ\ and \KS\ candidates 
is peaked towards
$|\cos\theta_d| = 1$ and the two-body decays of the spinless \DZ\ have a flat
distribution in $\cos\theta_d$. We require $\cos\theta_d$ to be in the
range $[-1.00,+0.95]$,
$[-0.875,+0.875]$ and $[-0.96,+0.96]$ 
for \KS\PZ, \PZ\PZ\ and \KS\KS\ final states, respectively.

\begin{figure}
\epsfig{file=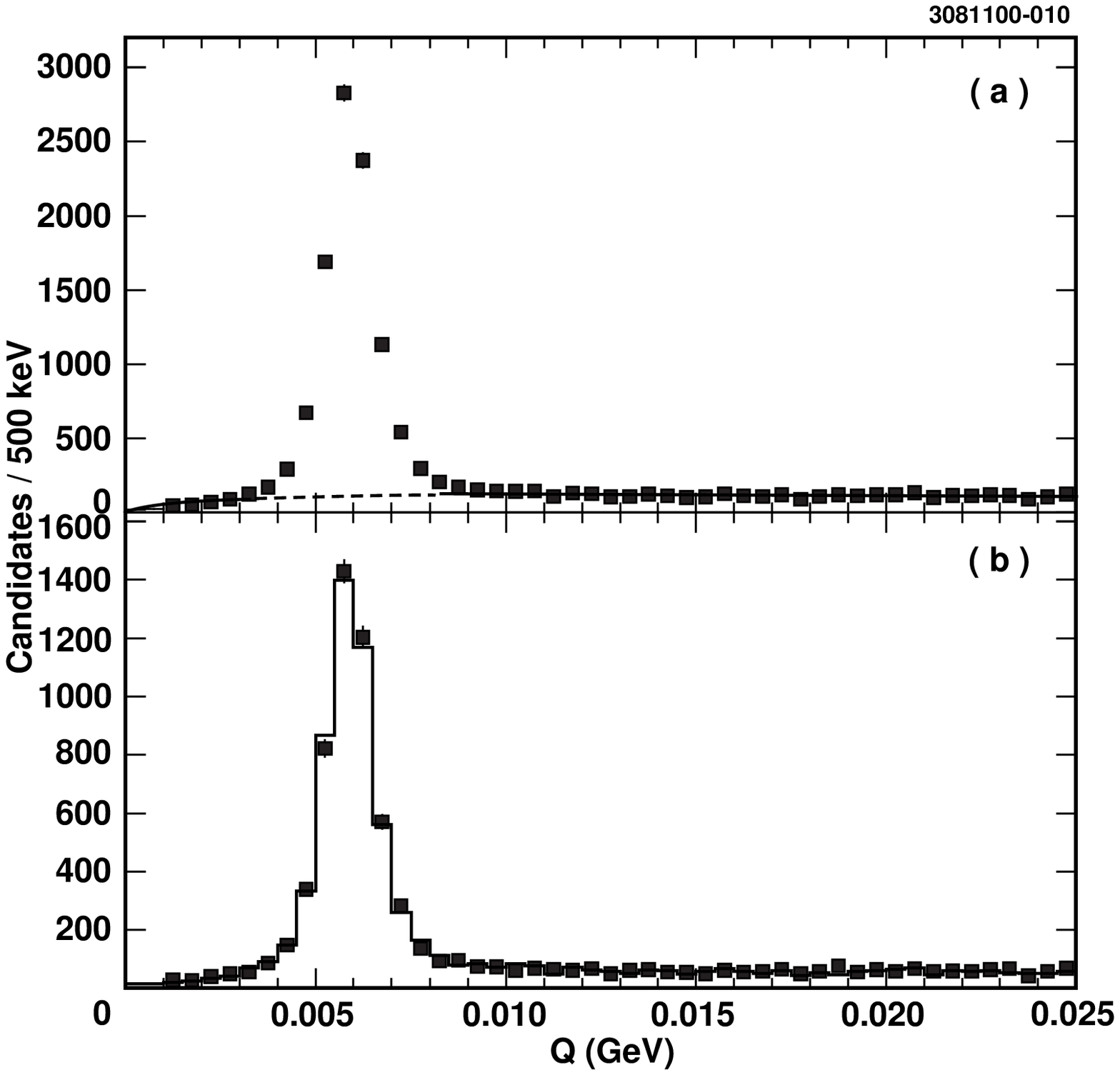,width=\linewidth}
\caption{
	(a) Fitted $Q$ distribution for $\DZ\to\KP$. 
	The points with error bars are the data, the 
	(barely visible) %v2
        solid line represents
	the fitted background and the dashed line shows the interpolation
	into the $Q$ signal region.
	(b) The $Q$ distributions for $\DZ\to\KP$ (points) and
	$\DZBAR\to\KP$ (histogram) candidates.
	}
\label{fig:kp}
\end{figure}

\begin{figure}
\epsfig{file=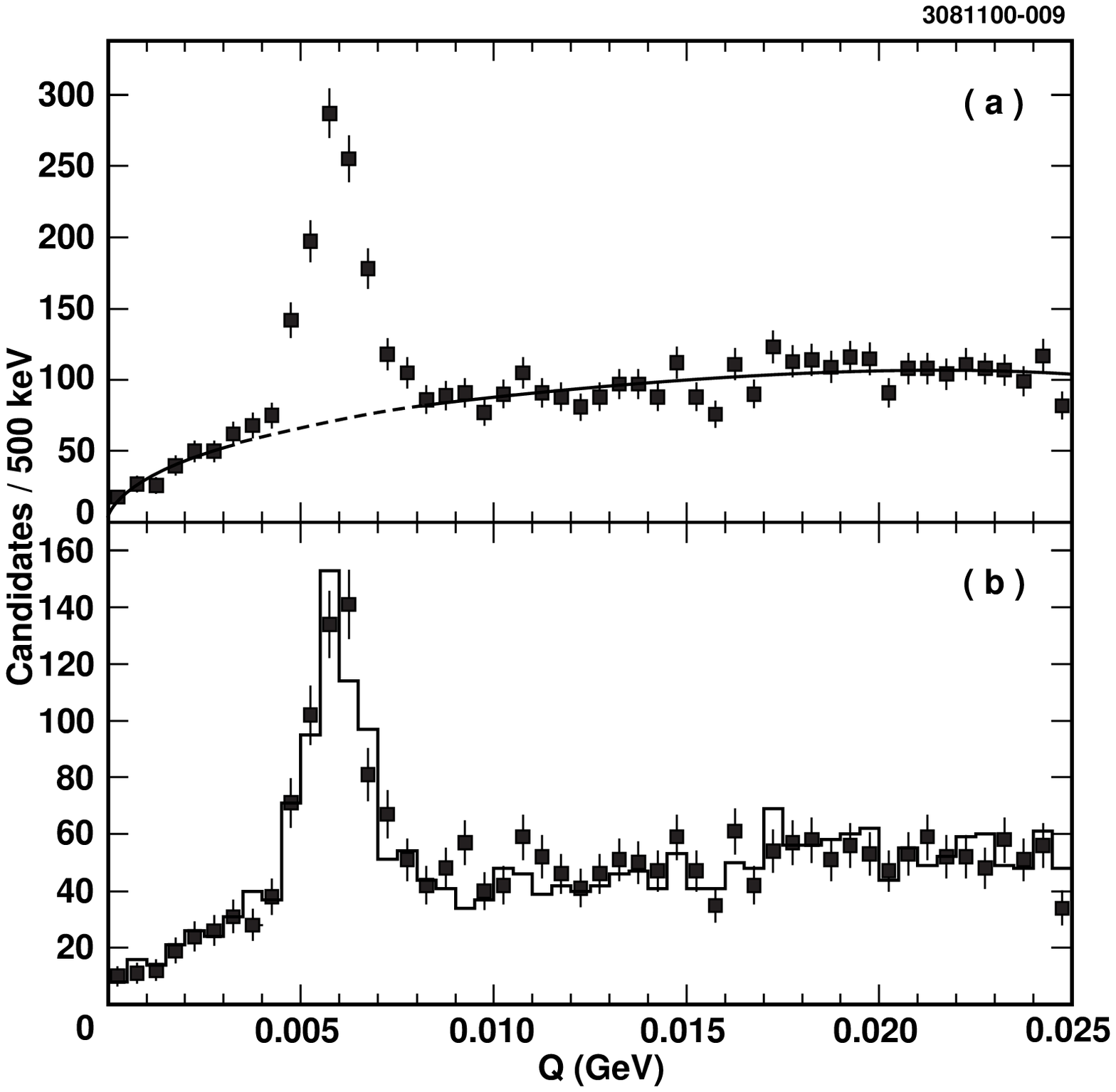,width=\linewidth}
\caption{
	(a) Fitted $Q$ distribution for $\DZ\to\PP$. 
	The points with error bars are the data, the solid line represents
	the fitted background and the dashed line shows the interpolation
	into the $Q$ signal region.
	(b) The $Q$ distributions for $\DZ\to\PP$ (points) and
	$\DZBAR\to\PP$ (histogram) candidates.
	}
\label{fig:pp}
\end{figure}

\begin{figure}
\epsfig{file=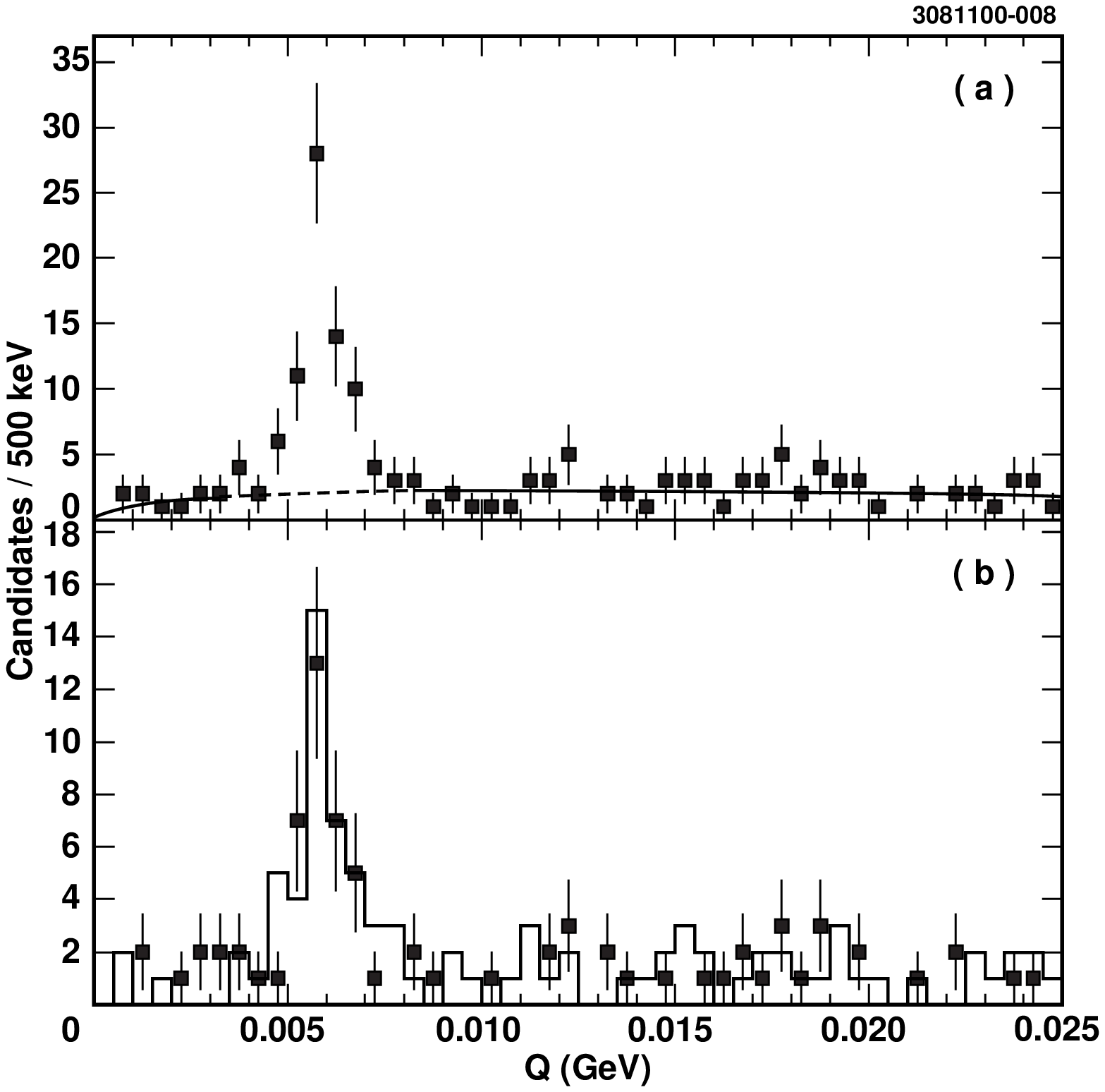,width=\linewidth}
\caption{
	(a) Fitted $Q$ distribution for $\DZ\to\KK$. 
	The points with error bars are the data, the solid line represents
	the fitted background and the dashed line shows the interpolation
	into the $Q$ signal region.
	(b) The $Q$ distributions for $\DZ\to\KK$ (points) and
	$\DZBAR\to\KK$ (histogram) candidates.
	}
\label{fig:kk}
\end{figure}

 \DZ\ candidates are selected by requiring $M$, the reconstructed \DZ\ candidate
mass, to be within 50, 65, and 18 MeV of the known \DZ\ mass~\cite{pdg} for
\KS\PZ, \PZ\PZ\ and \KS\KS\ final states, respectively. The $Q$ distributions
of the candidates in the three decay modes are shown in
Figures~\ref{fig:kp}, \ref{fig:pp} and \ref{fig:kk}, respectively, where
$Q$ is the energy release, $Q\equiv M(\DZ\slowpi) - M - M_{\pi^+}$,
$M(\DZ\slowpi)$ is the \DZ\SLOWPI\ invariant mass and $M_{\pi^+}$ is the
charged pion mass~\cite{pdg}. A prominent peak indicative 
of $\DSTP\to\DZ\SLOWPI$
decays in observed in all three distributions.

 The sum ${\cal S}$ of 
the number of %v2
\DZ\ and \DZBAR\ candidates  to a given final state 
(the denominator in Eqn~(\ref{eqnA}) ) is
determined by fitting the background in the $Q$ distribution.
The background shape is approximated as a non-relativistic threshold
function with 
first and second order % v2
relativistic corrections
$B(Q) = a Q^{1/2} + b Q^{3/2} + c Q^{5/2}$ and
the signal region \QSIG\ is excluded from the fit.
The interpolated background in the signal region is determined from
the fit and subtracted from the total number of \DZ\ and \DZBAR\
candidates to determine ${\cal S}$. For the three
decay modes under investigation, we obtain
% ${\cal S}(\KP) = 9098.8\pm 151.3$, 
% ${\cal S}(\PP) = 810.4\pm 88.5$ and
% ${\cal S}(\KK) = 65.0\pm 14.0$ 
% sigfigs
 ${\cal S}(\KP) = 9099\pm 151$, 
 ${\cal S}(\PP) = 810\pm 89$ and
 ${\cal S}(\KK) = 65\pm 14$ 
where the quoted uncertainty includes
the uncertainty due to the background interpolation.
The numerator in Eqn.~\ref{eqnA}
is determined from the difference in the number of \DZ\ and \DZBAR\ 
candidates in the region \QSIG. The measured raw asymmetries are
\RAWKP, \RAWPP\ and \RAWKK\  
for the \KP, \PP\ and \KK\ final states, respectively.
This method of determining the asymmetry implicitly assumes that the background
is symmetric. As shown in Figures~\ref{fig:kp}, \ref{fig:pp} 
and \ref{fig:kk}, the
$Q$ distributions are indeed symmetric outside the region \QSIG\ to the
statistical precision available.

  We have measured the 
momentum-dependent 
detector- or reconstruction-induced 
slow pion asymmetry by selecting charged pions from \KS\ decays using the
same selection criteria used to select \DSTP\ daughters. Since the inner
detector material differs for the two configurations, we take the measured 
asymmetry as a function of momentum for each configuration and weight it 
by the \SLOWPI\ 
spectrum from \DSTP\ decays. 
The asymmetry measured for CLEO II and CLEO II.V
is 
%$(-0.196\pm 0.335)\%$ and $(+0.183\pm 0.225)\%$, 
%sigfigs
$(-0.20\pm 0.34)\%$ and $(+0.18\pm 0.23)\%$, 
respectively, where the uncertainties are statistical only.
The overall asymmetry, weighted by the luminosity accumulated with each
configuration, is 
%$(+0.052\pm 0.187)\%$. 
%sigfigs
$(+0.05\pm 0.19)\%$. 
We apply a correction to the measured raw asymmetries
for \DZ\ decays for this small bias.

 We have also investigated the possibility 
of a bias in the fitting 
method by repeating the procedure on samples of simulated events.  % v2
$Q$ distributions of
\DZ\ and \DZBAR\ candidates with known asymmetries ranging between $\pm 50\%$
and with statistics comparable to that of the data were constructed from 
simulated events. 
%No bias was observed in the measured asymmetries of these samples. 
In addition we measured the asymmetry by fitting the $Q$ 
distributions 
to obtain the number of   %v2
\DZ\ and \DZBAR\ candidates separately, instead of the
procedure described above. We also used an alternative parameterization of the 
 background that was found to accurately model the background shape in 
simulated events: $B(Q) = e^{cQ} Q^n$ where $c$ and $n$ are fitted parameters. 
The 
magnitude of the %v2
average bias observed in these samples was $<0.1\%$ and we estimate 
the systematic uncertainty from the fitting procedure to be 0.5\% 
(approximately twice the RMS of the asymmetries
measured in the simulated samples).

 Finally we must take into account 
the possible effects of asymmetries in other
\DZ\ decay modes that have mistakenly been reconstructed in one of the three
selected decay modes. We can write the measured asymmetry ${\cal A}_m$ 
as ${\cal A}_m  = ({\cal A}_s + r{\cal A}_b)/(1+r)$ where 
${\cal A}_s$ is the asymmetry of the selected \DZ\ decay mode ("signal"),
${\cal A}_b$ is the asymmetry of other \DZ\ modes ("background") that
are reconstructed as the signal mode and $r$ is the ratio of the rates of 
the background to the signal. Background from decays of the form
$\DZ\to \KS\PZ\PZ$, $\pi^+\pi^-\PZ\PZ$, \PZ\PZ\PZ\ or \KS\KS\PZ\ can 
not contribute
because their reconstructed mass is approximately one pion mass below the \DZ\ 
mass and well outside the allowed range for $M$. Background from
$\DZ\to h^+h^-\PZ$  ($\DZ\to h^+h^-\KS$ and $\DZ\to h^+h^-h^+h^-$) 
can contribute to the reconstructed
\KP\  (\KK) yield when the  $h^+h^-$ mass is near the \KS\ mass.
There are no backgrounds of this sort that can contribute to the \PP\ yield.
The magnitude and asymmetry of this background can be directly measured
using $\KS\to \pi^+\pi^-$ candidates with $M(\pi^+\pi^-)$ in sidebands
either just below  or above  the $M(\pi^+\pi^-)$ mass range
for standard \KS\ candidate selection. Using the same analysis 
procedure for these
sideband candidates, 
we measure 
% $r(\KK) = 0.16\pm 0.11$ and ${\cal A}_b(\KK) = (+40.00\pm 41.69)\%$
% sigfigs
$r(\KK) = 0.16\pm 0.11$ and ${\cal A}_b(\KK) = (+40\pm 42)\%$
and 
%$r(\KP) = 0.03\pm 0.02$ and ${\cal A}_b(\KP) = (-5.49\pm 5.07)\% $ 
% sigfigs
$r(\KP) = 0.03\pm 0.02$ and ${\cal A}_b(\KP) = (-5.5\pm 5.1)\% $ 
where the uncertainties
are statistical only. The relative background rate for \KK\ is substantially
higher than that for \KP\ because the primary decay mode
contributing to the \KK\ background
is the Cabibbo-favored $\DZ\to\KS \pi^+\pi^-$ while the main contributors
to \KP\ are the Cabibbo-suppressed $\DZ\to\pi^+\pi^-\PZ$  and 
kinematically asymmetric $\DZ\to K^-\pi^+\PZ$ decays.

 Correcting the measured raw asymmetries for the slow pion reconstruction
bias and the rate and asymmetry of the background, we obtain
$\ACP(\KP) = \AKP$, $\ACP(\PP) = \APP$ and $\ACP(\KK) = \AKK$ where the
uncertainties contain the combined statistical and systematic uncertainties.
All systematic uncertainties, except for that assigned for possible %v2
bias in the fitting method, are determined from the data and would %v2
be reduced in future higher luminosity samples. %v2
All measured asymmetries are consistent with zero 
and no indication of significant \CPV\ is observed. The former 
measurement is a substantial improvement over the previous CLEO 
measurement~\cite{cleo.cpv} and supersedes it. This is the first measurement of
the asymmetry in \DZ\ decays to the Cabibbo-suppressed final states
\KK\ and \PP. 

%%% sept 2000 prl style acknowledgements with REU additions
 We gratefully acknowledge the effort of the CESR staff 
in providing us with
excellent luminosity and running conditions.
M. Gramaglia and L.D. Borum II 
thank the REU program of the National Science Foundation. 
M. Selen thanks the PFF program of the NSF and the Research Corporation, 
A.H. Mahmood thanks the Texas Advanced Research Program,
F. Blanc thanks the Swiss National Science Foundation, 
and E. von Toerne thanks the Alexander von Humboldt Stiftung for support.
This work was supported by the National Science Foundation, the
U.S. Department of Energy, and the Natural Sciences and Engineering Research 
Council of Canada.

\end{document}